# Theory of the center-of-mass diffusion and viscosity of microstructured and variable sequence copolymer liquids

Guang Shi[1] and Kenneth S. Schweizer[1-4*]

1. Department of Materials Science, University of Illinois, Urbana, Illinois 61801, United States
2. Materials Research Laboratory, University of Illinois, Urbana, Illinois 61801, United States
3. Department of Chemical and Biomolecular Engineering, University of Illinois, Urbana, Illinois 61801, United States
4. Department of Chemistry, University of Illinois, Urbana, Illinois 61801, United States

* Author to whom correspondence should be addressed: kschweiz@illinois.edu

**Abstract**

Biomolecular condensates formed through the phase separation of proteins and nucleic acids are widely observed, offering a fundamental means of organizing intracellular materials in a membrane-less fashion. Traditionally, these condensates have been regarded as homogeneous isotropic liquids. However, in analogy with some synthetic copolymer systems, our recent theoretical research has demonstrated that model biomolecular condensates can exhibit a microemulsion-like internal structure, contingent upon the specific sequence, inter-chain site-site interactions, and concentrated phase polymer density. In this study, we present a microscopic dynamical theory for the self-diffusion constant and viscosity of concentrated unentangled A/B regular multiblock copolymer solutions. Our approach integrates static equilibrium local and microdomain scale structural information obtained from PRISM integral equation theory and the time evolution of the autocorrelation function of monomer scale forces at the center-of-mass level that determine the polymer diffusion constant and viscosity in a weak caging regime far from a glass or gel transition. We focus on regular multi-block systems both for simplicity and for its relevance to synthetic macromolecular science. The impact of sequence and inter-chain attraction strength on the slowing down of copolymer mass transport and flow due to local clustering enhanced collisional friction and retardation of motion due to emergent microdomain scale ordering are established. Analytic analysis and metrics employed in the study of biomolecular condensates are employed to identify key order parameters that quantity how attractive forces, packing structure, multiblock sequence, and copolymer density determine dynamical slowing down above and below the crossover to a fluctuating polymeric microemulsion state.

1. **Introduction**

In the last decade, liquid-liquid phase separation (LLPS) has gained prominence as a pivotal cellular process [1–8]. Characterized by the formation of a dense coexisting phase called a biomolecular condensate, LLPS plays a vital role in various biological functions ranging from disease progression and cancer development [9–11], to gene regulation and responses [12–15], to external stressors [16–18]. To grasp the biological roles of these condensate-forming proteins and nucleic acids, a deep dive into their biophysical properties is imperative. These properties dictate not only the in-vivo thermodynamics of condensate formation, but also determine the dynamic interactions between the condensates and their surroundings, impacting molecular transport into and out of the condensate [19–21].

*In vitro* experiments often characterize biomolecular condensates as liquid droplets, given their ability to fuse upon contact and their global spherical shape [7,22]. However, the classification of a material as liquid hinges on the specific timescales of the dynamic processes involved. For instance, the timescales of fusion between two droplets can vary significantly across different condensate-forming biomolecules. Some proteins exhibit fusion within seconds [23], while others may not fully merge into a spherical shape even after several hours [6]. The viscosity for different biomolecular condensates have been found span three orders of magnitude as measured using the micropipette aspiration method [19,24]. Several studies have found that the reconstituted biomolecular condensates are not simple liquids but rather exhibit strongly viscoelastic properties [25–28]. Moreover, physical aging has been observed in some systems, where the morphology, diffusivity and viscosity of the condensates continuously evolves over time [5,18,29]. Additionally, the concept of a Maxwell glass has been proposed to describe the dynamic characteristics of some condensate-forming proteins [30,31].

The sticker–spacer model framework inspired by polymer physics [32] has been proposed as a useful minimal model for many of the condensate-forming systems. This is an A/B copolymer model where there are two types of interaction sites that are characterized by strongly attractive sticker-sticker interactions and repulsive or weakly attractive interactions between spacers and/or spacer-stickers. In our previous recent study [33], we have used liquid state theory (the polymer reference interaction site model, or PRISM, theory [34,35]) to investigate the thermodynamics, phase behavior and structure of such a copolymer model that is either of a regular (e.g., diblock, triblock) or an aperiodic nature germane to biology, in concentrated solutions and melts motivated by the condensate problem. Here we build on this advance to formulate a general microscopic dynamical theory for globally disordered A/B copolymer liquids by combining ideas of polymer, colloid, and liquid-state statistical mechanics at the level of segment-scale correlated space–time intermolecular forces to predict the center-of-mass diffusion constant and viscosity of such systems.

Our recent biophysically motivated study [33] revealed two distinct regimes in the phase diagram for a sticker-spacer model of biomolecular condensates. At low to intermediate copolymer volume fractions, the system undergoes macroscopic phase separation, akin to the putative liquid-liquid phase separation observed in biomolecular condensates. At higher monomer fractions, instead of phase separating into two phases, upon reducing temperature or increasing the sticker-sticker attraction the system continuously (but sharply) transitions to a microemulsion-like globally homogeneous state, characterized by strong local clustering and microdomain structuring beyond a system-specific and copolymer sequence dependent Lifshitz-like point $\phi > \phi_L$.

Our prior results [33] are relevant not only to biomolecular condensates, but also for synthetic copolymer systems with ordered and disordered sequences which are known to exhibit strong microdomain scale fluctuations [36] in the globally homogeneous state including of a polymeric microemulsions nature depending on system composition [36,37]. The focus of the present article is to address the question of how does the emergence of such polymeric microemulsion-like structure and local clustering impact dynamics on the macromolecular scale? Although we believe the proposed approach is valid for both periodic and aperiodic copolymer sequences, in this initial article we focus on implementation of the theoretical ideas for regular A/B multiblock copolymers of highly variable block size. We consider a multiblock model composed of the same A and B monomers at a fixed composition with only sequence (block length) varied. Such systems can be synthesized using the methods of polymer chemistry [36,38]. Hence our results are also relevant to synthetic macromolecular science, in addition to providing a foundation for treating biomolecular condensates in the future.

Our basic proposed physical picture is the following: the increase of monomer-scale contacts between sticker contacts due to local clustering and the emergence of larger length scale microdomain structure have distinct, but highly correlated, effects on polymer self-diffusion and bulk viscosity. We adopt the general mode coupling theory (MCT) idea [39] adapted to polymers [40–43] to separate the interchain force-force time correlation experienced by a tagged copolymer into a fast local component due to enhanced collision induced friction associated with very local clustering, and a slowly relaxing correlated dynamic density fluctuation many body component which has both local monomer scale cage and chain connectivity mediated microdomain scale components. Due to the underlying tendency of the copolymers to macro- or micro- phase separate, we argue (and provide evidence for) that typical copolymer systems are in a dynamical "weak caging" regime [44] far from the strong caging regime characterized by long lived glassy or gel (due to physical bonds) like transient localization. In addition, since the typical sequence lengths of model biomolecules studied [20,45–48] are relatively short ($N < 300$), for the concentrated solutions of typical relevance one expects that topological entanglement effects [49] are not important. These considerations suggest a simplifying separation of time scales exists for many systems corresponding to the macromolecular relaxation or diffusion time of single copolymers being long compared to the time scale of the relaxation of interchain force time correlations that slow their dynamics, the definition of a "weak caging" regime [44]. In this situation, the effect of local clustering and emergent microdomains on tagged polymer dynamics should be well captured by a non-self-consistent version of mode coupling theory (nsc-MCT [50–52]), which has been found to be quantitatively successful in predicting experimental data in systems such as concentrated hard sphere and repulsive colloid suspensions [50–52] and ring polymer liquids [44] in the initial slowing down regime that proceeds the crossover to even slower activated relaxation associated with strong transient localization in glass and gel forming systems [39,50].

Using the developed theory, we systematically investigate the sticker/spacer copolymer dynamics for a foundational model of regular multi-block copolymers with a composition of 50% sticky monomers of fixed degree of polymerization but highly variable block size. The effect of varying sticker-sticker attraction strength and block size on slowing down of center-of-mass diffusion and collective viscosity is systematically investigated, with a particular focus on the deviations from simple Rouse-like dynamics [53] due to inter-chain dynamical effects associated with emergent microdomains. Our theory is not limited to regular sequences and can be applied to irregular sequences of synthetic and biological origin.

Section 2 describes the polymer model employed, briefly reviews PRISM theory, and formulates a non-self-consistent copolymer mode coupling theory and the computation of the collective component of the shear viscosity. Sections 3, 4, and 5 present our main results, including for the homopolymer system, copolymer fluids, and results obtained using simplified PRISM-RPA theory [33,54] that ignores the important fluctuation and correlation effects on structure and dynamics. Finally, we summarize our key findings and discuss the limitations and future directions of the present study in section 6. Various technical details, a brief summary of selected elements of the background theory, and additional results that support the conclusions drawn in the main article are collected in the Supplementary Information (SI).

## 2. Model and Theories

### 2.1 Model

Our focus is on concentrated copolymer solutions (modeled using implicit solvent) and melts where conformations are close to ideal random coils, which is also supported by the experimental observations for biomolecular condensates [55,56]. In this concentrated regime, the importance of solvent-induced hydrodynamic interactions are also minimized and are not addressed here. We consider regular block copolymers (Fig. 1a) for simplicity, but the approach can be applied to aperiodic copolymers of biological and synthetic interest.

The selection of A-A, A-B, and B-B interactions follows our previous work [33] that studied the phase behavior and low wavevector (microdomain scale) structure of the sticker-spacer model for biomolecular condensate formation. An ideal freely-jointed chain (FJC) model is adopted that consists of spherical interaction sites of type A and B of identical hard core diameters, $\sigma$. The inter-chain interactions between A-B and A-A sites are taken to be purely hard-core, with B-B interactions including an effective short-range attraction of an exponential functional form:

$$u_{\text{BB}}(r) = \begin{cases} -\epsilon_{\text{BB}} \exp(-(r-\sigma)/\alpha) & \text{if } r \geq \sigma \\ \infty & \text{if } r < \sigma \end{cases}$$

Here, $\alpha$ defines the attraction range, and $\epsilon_{\text{BB}}$ is the attraction energy at contact (Fig. 1(a)). The model control parameters are chain length $N$, the reduced temperature $k_B T/\epsilon_{\text{BB}}$, the total polymer packing fraction $\phi = \pi \rho \sigma^3/6$, where $\rho$ is the total site number density, and chain persistence length taken to equal the bond length of the FJC, $l = (4/3)\sigma$ [35,57]. Without loss of generality, the length and energy scales are reported in units of $\sigma$ and $k_B T$, respectively. We fix $N = 256$, which is comparable to the size of many intrinsically disordered proteins [20,45–48], and is plausibly short enough that entanglement effects are not present [53].

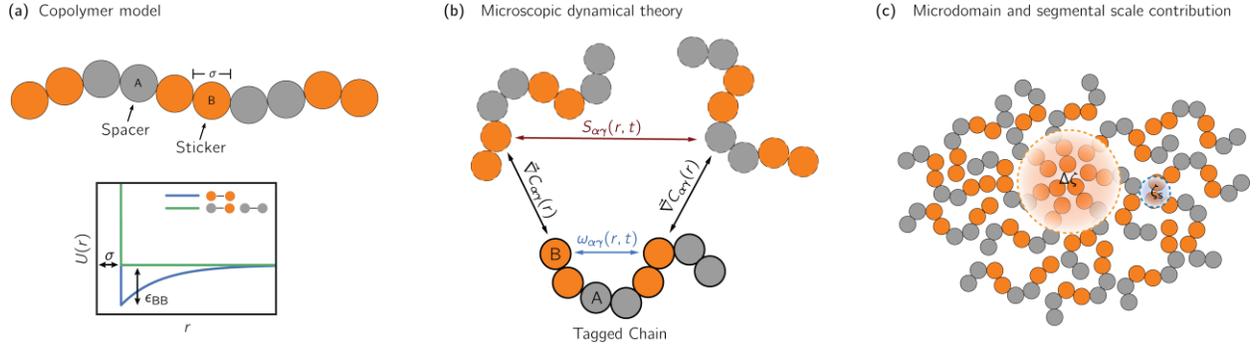

*Figure 1: (a) Multiblock copolymer model investigated in this work. The B–B (sticker–sticker) sites attract via an exponential potential plus a hardcore repulsion, and A–A/A–B sites interact solely as hard spheres. (b) The dynamical theory considers the interchain force-force autocorrelation (memory) function on a tagged chain exerted by all the surrounding copolymers*

*and involves effective site-site interchain forces, $k_B T \vec{\nabla} C_{\alpha\gamma}(r)$. The force time correlations on two different monomers of a tagged chain relax dynamically via tagged chain motion encoded in the intramolecular dynamical correlation, $\omega_{\alpha\gamma}(r,t)$ and the collective dynamical density fluctuations of surrounding chains encoded via $S_{\alpha\gamma}(r,t)$. (c) Schematic of the contribution to the total center-of-mass friction from microdomain scale clustering of attractive stickers ($\Delta\zeta$) and that from the short-time local segmental scale contact collision between monomers ($\zeta_s$).*

**2.2 PRISM theory**

Equilibrium structural correlation information is determined from PRISM integral equation theory [34,35,57] and employed as input to construct our dynamical theory for globally homogeneous, isotropic, single-phase copolymer fluids. PRISM theory has been successfully applied to study regular copolymer melts and solutions [58–61], regular associating copolymers [43,62,63], macroscopic demixing in binary polymer blends [64], block copolymers in solutions [65,66], and most recently model biomolecular condensates [33].

PRISM theory relates site-site intermolecular pair correlations, intramolecular pair correlations (statistical conformation), and direct correlation functions based on the generalized Ornstein-Zernike (or Chandler-Andersen) equation written in Fourier space (wavevector,$k$) [57],

$$\mathbf{H}(k) = \mathbf{\Omega}(k)\mathbf{C}(k)[\mathbf{\Omega}(k) + \mathbf{H}(k)] \qquad (1)$$

Here, $H_{\alpha\gamma}(k) = \rho_{\alpha\gamma}^{\text{pair}}(k)h_{\alpha\gamma}(k)$ is the Fourier transform of the dimensional total intermolecular site-site correlation function between species $\alpha$ and $\gamma$, $\rho_{\alpha\gamma}^{\text{pair}} = \rho_\alpha \rho_\gamma$ with $\rho_\alpha$ and $\rho_\gamma$ the number density of species $\alpha$ and $\gamma$, $\Omega_{\alpha\gamma}(k) = \rho_{\alpha\gamma}^{\text{site}}\omega_{\alpha\gamma}(k)$ is the dimensional intramolecular correlation function, and $C_{\alpha\gamma}(k) = c_{\alpha\gamma}(k)$ is the direct correlation functions that describes effective or renormalized intermolecular site-site pair interactions. Notationally, $\rho_{\alpha\gamma}^{site} = \rho_\alpha + \rho_\gamma$ if $\alpha \neq \gamma$ and otherwise $\rho_{\alpha\gamma}^{site} = \rho_\alpha$, and $\omega_{\alpha\gamma}(k)$ is the copolymer sequence-dependent intramolecular correlation function (see SI for the detailed derivation and mathematical expressions). The dimensional collective density fluctuation structure factor matrix is,

$$\tilde{\mathbf{S}}(k) = \mathbf{\Omega}(k) + \mathbf{H}(k) = [\mathbf{I} - \mathbf{\Omega}(k)\mathbf{C}(k)]^{-1}\mathbf{\Omega}(k) \qquad (2)$$

The entries of $\tilde{\mathbf{S}}(k)$ are the partial collective static structure factors $\tilde{S}_{\alpha\gamma}(k)$. In dimensionless form they are given by $S_{\alpha\gamma}(k) = (1/\rho_{\alpha\gamma})\tilde{S}_{\alpha\gamma}(k)$, where $\rho_{\alpha\gamma} = \sqrt{\rho_\alpha \rho_\gamma}$.

Following our previous work on biomolecular condensates [33] and for the same reasons explained there, the coupled PRISM equations are closed by the mean spherical approximation (MSA) [57,67] $c_{\alpha\gamma}(r) = -\beta u_{\alpha\gamma}(r)$ for $r > \sigma$, where $u_{\alpha\gamma}(r)$ is the non-contact tail potential between species $\alpha$ and $\gamma$. Eq. (1) are numerically solved using the Newton-Krylov method built in the pyPRISM package [68] with a choice of 32,768 discretized points in real and reciprocal space following previous studies [42].

**2.3 Copolymer non-self-consistent mode coupling theory for center of mass diffusion**

Our goal is to formulate a microscopic theory at the level of segmental scale repulsive and attractive forces for the copolymer center-of-mass diffusion constant whereby kinetic constraints on a tagged polymer are related to the ensemble-averaged pair correlation functions. For context, we first briefly explain the basic ideas adopted for the simpler system of a hard sphere colloidal suspension characterized by a packing fraction $\phi$ [39,50–52]. The focus is on high enough packing fractions (above approximately $\phi > 0.2$) that a particle or colloid has enough neighbors to be "caged" to some extent which significantly slows its motion, but not high enough (typically $\phi < 0.50 - 0.55$) to enter a transient localization regime characteristic of strong caging which drive a crossover to much slower glassy dynamics and ultimately kinetic vitrification [39,40,69]. In this regime, the friction experienced by a diffusing particle involves "weak caging" constraints and at short times independent (uncorrelated) binary collisions between a pair of hard spheres. Activated dynamics is not present or important, and experiments [70,71] and simulations find [72] the self-diffusion coefficient $D$ decreases by roughly a factor of ~10-30 as the packing fraction increases. In essence, in the weak caging regime the intermolecular force time correlation that impede tagged particle motion relax faster than the rate of mass transport. This time-scale separation results in an almost exponential one-step decay of time correlation function of single and collective particle dynamic structure factors, and an approximately Fickian mean squared displacement (MSD). Crucial for our present work, in the weak caging regime the physics is well captured by a non-self-consistent version of mode coupling theory (nsc-MCT) [44,50–52].

The above scenario breaks down at higher packing fractions where particles become transiently localized, the diffusion coefficient $D$ decreases much faster with increasing packing fraction, particle motion becomes intermittent, activated and highly non-Gaussian, and a two-step decay of dynamic structure factors and other time correlation functions emerges [39,69,73]. Such "glassy dynamics" results in a practical (laboratory) kinetic vitrification of dense colloidal suspensions at a packing fraction of ~0.58-0.6. To theoretically capture this requires a self-consistent MCT [39] and a beyond MCT treatment of dynamics including thermally activated hopping motion [50,74,75].

While it can be challenging to measure the packing fraction of biomolecular condensates, most experimental estimates place it within an intermediate regime of $\phi < 0.4$ [56,76–79], and hence in what we call the weak caging regime. This motivates our present generalization of weak caging theories to concentrated copolymer and model biomolecular condensate fluids, as successfully recently done in the different context of dense ring polymer melts and concentrated solutions [44].

The polymeric weak caging theory of the center-of-mass (COM) friction coefficient for a tagged chain is formally related to the autocorrelation function of the total forces exerted on all monomers of a tagged chain by the surrounding polymers [44],

$$\zeta_{CM} = \frac{\beta^{-1}}{3} \int_0^\infty dt \langle \mathbf{F}_{CM}(0) \cdot \mathbf{F}_{CM}(t) \rangle = \frac{\beta^{-1}}{3} \sum_{i,j=1}^N \langle \mathbf{f}_i(0) \mathbf{f}_j(t) \rangle \qquad (4)$$

where $\mathbf{F}_{CM}$ is the total intermolecular force exerted on a tagged chain by all surrounding polymers and $\mathbf{f}_i$ is the force experienced by a single segment $i$. The total COM friction constant is formally decomposed into two additive contributions: short-time friction for individual monomers due to the highly local and rapidly relaxing forces component, and a longer range and more slowing relaxation collective part associated with structural correlations beyond the local segmental scale.

We adopt the standard polymer physics perspective whereby the simple Rouse model [53] captures the rapid and spatially uncorrelated part of the local intermolecular force time correlations, and use polymeric nsc-MCT to determine the longer range slower relaxing component on beyond segmental length scales up to the macromolecular and (for copolymers) fluctuating microdomain scales. The latter contribution contains "nonlocal" contributions in the sense of structural packing mediated cross-correlations between forces on different segments of a tagged polymer (as signified by the off-diagonal terms in Eq.(4)), which depend explicitly on chain connectivity and copolymer sequence. The resulting total friction constant can thus be written as,

$$\zeta_{CM} = N\zeta_s + \Delta\zeta_{CM} = N\zeta_s + N\Delta\zeta = N(\zeta_s + \Delta\zeta) \qquad (5)$$

where $N$ is the chain length. The simple Rouse model contribution is $N\zeta_s$ where $\zeta_s$ is the short-time local segmental friction coefficient, which sets the elementary friction and time scale of the slower dynamics. Given its highly local nature, and the fact we are studying only the weak caging regime, we model $\zeta_s$ explicitly as due to independent repulsive binary collisions between segments which induces a local friction proportional to the inter-segment collision rate amplified by the tendency of attractive intermolecular interactions to enhance the probability of two sites on different polymers to contact [50,51]. The latter information enters via the contact values of the interchain pair correlation functions predicted by PRISM theory, $\zeta_s \propto \overline{g(\sigma)} = f_A g_{AA}(\sigma) + f_B g_{BB}(\sigma)$, where $g_{AA}(\sigma)$, $g_{BB}(\sigma)$ are the contact values for the A-A and BB site-site pair correlation function, respectively. Since in this study we report results with respect to a certain reference system, the prefactor does not matter, as explained below. The neglect of the explicit effect of attractive forces on the short-time friction is an approximation, albeit one that is reasonable if the attractive forces are not too large relative to thermal energy or spatially rapidly varying, as discussed in prior studies [80,81], and as we shall verify below to hold in our present work for the copolymer model adopted. This short-time segmental friction coefficient also enters the collective friction calculation in the weak caging theory since it sets a time scale for larger scale force relaxation as discussed below.

Now, $\Delta\zeta_{CM}$ in Eq. (5) is the collective or weak caging component of the friction coefficient, and $\Delta\zeta$ is the per-monomer contribution. Adopting the ideas of nsc-MCT [44,50–52] that many body dynamical caging effects are controlled by products of the single polymer and collective density fields (real forces are projected onto effective forces determined by structural correlations) and extending it to the copolymer system, $\Delta\zeta_{CM}$ can be related to the static pair correlations functions and written in Fourier space (after factorizing 3 and 4 point dynamical correlations into products of pair dynamic correlation functions) as,

$$\Delta\zeta_{CM} = \frac{\beta^{-1}}{3}\int_0^\infty dt \sum_{\alpha,\gamma} \int \frac{d\mathbf{k}}{(2\pi)^3} \mathbf{k}^2 \widetilde{\omega}_{\alpha\gamma}(k,t)[\mathbf{C}\cdot\widetilde{\mathbf{S}}(k,t)\cdot\mathbf{C}]_{\alpha\gamma} \qquad (6)$$

where $\mathbf{C}$ is the matrix of direct correlation functions, and $\widetilde{\omega}_{\alpha\gamma}(k,t)$ and $\widetilde{\mathbf{S}}(k,t)$ are the time-dependent intramolecular correlation function matrix and dynamic collective partial structural factor, respectively. The indices $\alpha$ and $\gamma$ denote the monomer types, A and B. Note that both $\widetilde{\omega}_{\alpha\gamma}(k,t)$ and $\widetilde{\mathbf{S}}(k,t)$ are not normalized, i.e. they contain stoichiometric factors, and $\widetilde{\omega}_{\alpha\gamma}(k,t) = n_{\alpha\gamma}\omega_{\alpha\gamma}(k,t)$ where $n_{\alpha\gamma} = n_\alpha$ if $\alpha = \gamma$ and $n_{\alpha\gamma} = n_\alpha + n_\gamma$ The collective partial dynamic structure factors $\widetilde{S}_{\alpha\gamma}(k,t) = \rho_{\alpha\gamma}S_{\alpha\gamma}(k,t)$, where $\rho_{\alpha\gamma} = \rho_\alpha$ if $\alpha = \gamma$ and $\rho_{\alpha\gamma} = \sqrt{\rho_\alpha\rho_\gamma}$ if otherwise. Here, $\omega_{\alpha\gamma}(k,t)$ and $S_{\alpha\gamma}(k,t)$ are dimensionless.

There are three distinct force correlation pathways in Eq. (6), spacer-spacer, sticker-spacer, and sticker-sticker. We physically expect (and verify below) that the sticker-sticker pathway is dominant at the lower reduced temperatures of primary interest where a fluctuating microemulsion type of structure emerges in the copolymer liquid. The real-space interpretation of Eq. (6) is shown schematically in Fig.1(b). Forces between two sites of species $\alpha$ and $\gamma$ is $\mathbf{f}(r) = k_BT\vec{\nabla}C_{\alpha\gamma}(r)$ where $C_{\alpha\gamma}(r)$ accounts both excluded volume constraints and the sticky-sticky monomer attraction at the segmental scale. The copolymer systems investigated in this study exhibit strong clustering and fluctuating microdomains. The decomposition of the total friction into two contributions in Eq. (5) provides a natural interpretation of the different contributions to the COM diffusion constant of a tagged chain due to clustering on the microdomain length scale and the short-time segmental-scale binary collisions between monomers, as illustrated in Fig. 1(c).

To solve Eqs. (4)-(6) requires $\widetilde{\omega}_{\alpha\gamma}(k,t)$ and $\widetilde{\mathbf{S}}(k,t)$ which are intractable since they contain the full many body dynamics. In the weak caging regime spirit, we adopt a dynamically effective homopolymer liquid approximation corresponding to the well-known Markovian dynamic RPA expression for $\widetilde{\omega}_{\alpha\gamma}(k,t)$ [41,53,82],

$$\widetilde{\omega}_{\alpha\gamma}(k,t) = \widetilde{\omega}_{\alpha\gamma}(k)\exp\left(-k^2 D_s t/\omega(k)\right) \qquad (7)$$

where $\omega(k)$ and $\widetilde{\omega}_{\alpha\gamma}(k)$ are the total and partial intramolecular correlation functions, respectively, and $D_s$ is short-time local segmental diffusion constant, $D_s = k_B T/\zeta_s$ where $\zeta_s$ is the local segmental friction constant. For the collective dynamic structure factors $\widetilde{\mathbf{S}}(k,t)$, we adopt the analogous dynamic RPA expression which obey the diffusive evolution equation [39,40,42,43,83],

$$\frac{\partial}{\partial t}\mathbf{S}(k,t) = -k^2 \mathbf{H}(k)\mathbf{S}^{-1}(k)\mathbf{S}(k,t) \qquad (8)$$

For simplicity, and because our focus is only the COM diffusion constant which averages over the friction on all A and B monomers of the copolymer chain, we adopt $D_A = D_B \equiv D_s = k_B T/\zeta_s$ where for a copolymer $\zeta_s$ is the chain-averaged *average* friction constant discussed above. Hence, one has $H_{11} = H_{22} = D_s$ and $H_{12} = H_{21} = 0$ in Eq. (8). The solutions of Eq. (8) can be written in the form of $S_{\alpha\gamma}(k,t) = a_{\alpha\gamma}\exp(-\Lambda_I t) + b_{\alpha\gamma}\exp(-\Lambda_C t)$, with the explicit details given in the SI. The idea of using a dynamic RPA equation for intra- and inter-chain correlations that enter the autocorrelation of the total force on a tagged polymer chain is based on the central physical picture of the present work that the forces on a tagged polymer relax faster than COM polymer diffusion, the "weak caging" regime. Effectively, force relaxation dynamics obey a diffusive, length-scale-dependent (relaxation times depend on wavevector) Rouse-like description. Eq. 6 is calculated using Eqs. (7)-(8), and the full expression and derivation is given by Eq. 9 in the SI. As a relevant limit, for the homopolymer fluid all partial direct correlation functions are identical, $C_{AA}(k) = C_{BB}(k) = C_{AB}(k) \equiv C(k)$, leading to $\mathbf{C} \cdot \tilde{\mathbf{S}}(k,t) \cdot \mathbf{C} = C^2(k)\left(\tilde{S}_{AA}(k,t) + \tilde{S}_{BB}(k,t) + 2\tilde{S}_{AB}(k,t)\right) = C^2(k)\tilde{S}_{tot}(k) = \rho C^2(k) S(k,t)$, where $S(k,t)$ is the homopolymer collective dynamic structure factor and $\rho$ is the total segmental number density. Eq. (6) then simplifies to,

$$\Delta \zeta_{HP} = \frac{\beta^{-1}}{3} \int_0^\infty dt \int \frac{d\mathbf{k}}{(2\pi)^3} \mathbf{k}^2 \rho C^2(k) S(k,t) \sum_{\alpha,\gamma} \tilde{\omega}_{\alpha\gamma}(k,t)$$
$$= \frac{N\rho\beta^{-1}}{3} \int_0^\infty dt \int \frac{d\mathbf{k}}{(2\pi)^3} \mathbf{k}^2 C^2(k) S(k,t) \omega(k,t) \quad (9)$$

where the second equality follows from $\sum_{\alpha,\gamma} \tilde{\omega}_{\alpha\gamma}(k,t) \equiv N\omega(k,t)$. $\omega(k,t)$, and $S(k,t)$ follows the dynamic random-phase approximation, $\omega(k,t) = \omega(k)\exp(-k^2 D_s t/\omega(k))$ and $S(k,t) = S(k)\exp(-k^2 D_s t/S(k))$. Eq. (9) is identical to the collective friction expression for a homopolymer fluid, which below we adopt as an athermal reference system where interactions are purely hard core.

**2.4 Single Chain Entropic and Collective Viscosities**

In general for polymeric systems, there are two different types of stress and hence viscosities: a "collective" viscosity associated with the autocorrelation of all stresses (including interchain force contributions) as present in all forms of matter [41,67,84], and the conformationally flexible polymer specific contribution associated with intrachain entropic stresses determined by chain connectivity considerations [53]. The latter is the focus of the Rouse model of polymer physics [53], and for unentangled homopolymer melts and concentrated solutions is assumed to be dominant since it is N-dependent in contrast (per standard assumptions) to its collective analog. It is given by [53]

$$\eta_R = \frac{1}{6}\frac{\rho}{N\beta}\frac{R_g^2}{D_{CM}} = \frac{1}{6}\rho R_g^2 \zeta_s(1 + \Delta\zeta/\zeta_s) \propto \rho N \zeta_s(1 + \Delta\zeta/\zeta_s) \quad (10)$$

where we use the subscript $R$ to denote $\eta_R$ is the effective single-chain entropic contribution to the total viscosity, per the Rouse model description.

The collective viscosity contribution follows from the time-dependent shear modulus as,

$$G(t) = \frac{1}{k_B T V}\langle \sigma^{xy}(0) e^{\Omega t} \sigma^{xy}(t)\rangle \quad (11)$$

where $\sigma^{xy}$ is the total microscopic stress tensor,

$$\sigma^{xy} = \frac{1}{2}\sum_{i,j}^{N} R_{ij}^x \frac{\partial u_{ij}(\mathbf{R}_{ij})}{\partial R_{ij}^y} \quad (12)$$

and $R_{ij}^x$ is the $x$ component of the displacement vector between sites $i$ and $j$. The summation is over all sites of the system and $e^{\Omega t}$ is the time evolution operator. Following literature studies [42,83,85], the collective viscosity for polymeric mixture systems (applicable to our copolymer system) the corresponding collective viscosity follows from a time integration as,

$$\begin{aligned}\Delta\eta &= \int_0^\infty dt\, G(t) = \frac{k_B T}{60\pi^2}\int_0^\infty dt \int_0^\infty dk\, k^4 \text{Tr}[\mathbf{V}^2(k,t)] \\ \mathbf{V}(k) &= \frac{d\mathbf{S}(k)}{dk}\mathbf{S}^{-1}(k)\mathbf{S}(k,t)\mathbf{S}^{-1}(k)\end{aligned} \quad (13)$$

where Tr is the trace operator. The complete expression for $\Delta\eta$ is,

$$\Delta\eta = \frac{k_B T}{60\pi^2} \int_0^\infty dt \int_0^\infty dk \sum_{\alpha,\beta,\gamma,\delta}^2 \sum_{\alpha',\beta',\gamma',\delta'}^2 \frac{dS_{\alpha\beta}(k)}{dk} \frac{dS_{\alpha'\beta'}(k)}{dk} S_{\alpha\delta}^{-1}(k) S_{\alpha'\delta'}^{-1}(k) S_{\beta\gamma}^{-1}(k) S_{\beta'\gamma'}^{-1}(k) S_{\gamma\gamma'}^{-1}(k,t) S_{\delta\delta'}^{-1}(k,t)$$

(14)

For the AB copolymer model investigated, there 256 terms that enter Eq. (14) associated with the diagonal and cross correlations of stresses. Eq. (13) is derived using standard MCT concepts [83] where stresses are projected onto the slow collective density fluctuation bilinear variables, and 4-point correlations are factorized into products of pair correlations. The $\Delta\eta$ calculated from Eq. (14) correctly diverges at a critical point (or spinodal) due to the presence of a diverging static correlation length, which is germane to mean field theories of copolymer microphase separation.

Combining the single-chain and collective contributions leads to the total viscosity, $\eta = \eta_R + \Delta\eta$. Our primary focus is the relative contribution of the collective viscosity to the total viscosity, as represented by $\Delta\eta/\eta$. If $\Delta\eta/\eta$ is small, the classic (Rouse model) inverse linear relation between the total viscosity $\eta$ and the total COM diffusion $\zeta_{CM}$ still holds. However, if $\Delta\eta/\eta$ is large, we anticipate a deviation from the simple inverse linear relation, reflecting larger length scale correlations (e.g., emergent microdomains in a fluctuating microemulsion structure) as a source of stress storage and resistance to flow.

### 3. Homopolymer reference and structural characteristics for copolymer system

#### 3.1 Athermal homopolymer reference system

We first investigate the copolymer volume fraction, $\phi$, dependence of the total COM friction $\zeta_{CM}$ of the reference athermal (hard core) homopolymer fluid with $N = 256$. Fig.2(a) shows this quantity, normalized to its value at $\phi \to 0$, increases by two orders of magnitude as $\phi$ grows from near zero to $0.6$, and in a manner that is exponential for the concentrated regime of interest, $0.2 < \phi < 0.6$. We note that prior work [42,43] for same homopolymer FJC model found the ideal polymeric MCT glass transition occurs at $\phi \sim 0.6$, beyond which there is a crossover to activated dynamics. Hence, our present weak caging theory study that extends at most up to a packing fraction $0.6$ is qualitatively justified. Recall that the total COM friction constant $\zeta_{CM}$ has contributions from single-chain and collective parts. The former is the Rouse contribution $N\zeta_s$ and the latter is $N\Delta\zeta$ in Eq. (5), where the local segmental friction constant $\zeta_s$, proportional to the average contact value and linearly enters both contributions. The normalized $N\zeta_s(\phi)/N\zeta_s(\phi \to 0)$ is also shown in Fig.2 for comparison with the total friction. One sees that the friction constant grows more strongly compared to when only the Rouse single-chain contribution is included. In addition, it appears that the exponential growth of friction as a function of $\phi$ is only apparent if collective friction is included. Nevertheless, the full friction constant that includes the collective contribution differs from the classic single-chain Rouse result by at most a factor of two, suggesting the collective friction is basically a perturbative correction, per textbook arguments [53].

We now ask whether the inclusion of collective friction alters the $N$-dependence of total friction $\zeta_{CM}$ from the Rouse model prediction of a linear dependence on $N$. We find it does not. This can be analytically seen from the following scaling analysis. Consider the so-called dynamic vertex that enters in Eq. (9), $V(k) = k^2 C^2(k)(S^2(k)\omega^2(k))/(S(k) + \omega(k))$. Fig.2(b) shows $V(kR_g)$ as a function of non-dimensionalized wavevector $kR_g$ for $N = 256$ at different polymer packing fractions $\phi \geq 0.3$. For intermediate values of $\phi \sim 0.3$, the vertex is dominated by contributions for wavevectors $kR_g \sim O(1)$. For $kR_g \sim O(1)$, $\omega(kR_g \sim O(1)) \propto N$ and $C(kR_g \sim O(1))$ and $S(kR_g \sim O(1))$ are $N$-independent. This leads to a total collective friction dominated by the peak at $kR_g \sim O(1)$ of $V(k)$, with $\Delta\zeta_{CM} \propto NV(kR_g \sim O(1)) \propto (1/R_g)^2 N^2 \propto N$. On the other hand, for large $\phi$ the vertex is dominated by its cage scale local packing peak at $k\sigma \sim 6$, the position of which is $N$-independent. Moreover, $C(k)$ and $S(k)$ are both $N$-independent to leading order. As a result, $\Delta\zeta_{CM} \propto NV(k\sigma \sim 6) \propto N$. Since both the collective and single-chain friction contributions are linear in $N$, the total COM friction constant scales linearly with $N$. These analytic arguments are consistent with our full numerical calculations which show that $\zeta_{CM}$ scales linearly with $N$ for both $\phi = 0.3$ and $\phi = 0.6$ (Fig.2(c)). Thus, the weak caging approach does not modify the classic $N$ scaling predictions for $D$ and the single chain stress based viscosity of the Rouse model for liquids of dense ideal unentangled chains. Rather, the collective friction, though not negligible in general, simply renormalizes the short time friction constant that enters the classic Rouse theory.

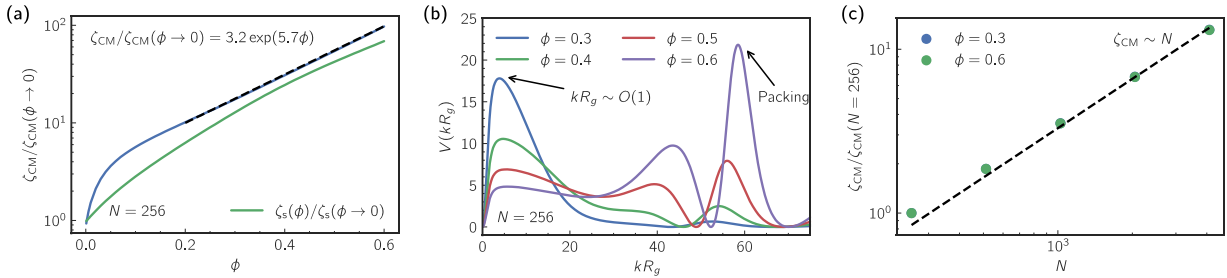

Figure 2: (a) Total center-of-mass friction constant $\zeta_{CM}$ as a function of $\phi$ for homopolymer fluids with a chain length $N = 256$, normalized by its value at $\phi = 0$. An exponential guide to the eye

*is shown. Green curve: the normalized growth of the single-chain contribution of friction.* ***(b)** Dynamic vertex $V(kR_g)$ as a function of non-dimensionalized wavevector $kR_g$ for $\phi = 0.3, 0.4, 0.5, 0.6$ and N=256. One sees a macromolecular scale peak at $kR_g \sim O(1)$ and a local cage or packing peak are predicted; note that $R_g = \sqrt{1/6}lN^{1/2} \approx 8.7\sigma$ with $l = (4/3)\sigma$.* ***(c)** N-dependence of total center-of-mass friction $\zeta_{CM}$ for the homopolymer fluid at $\phi = 0.3$ and $\phi = 0.6$. The data points are obtained from calculations for $N = 256, 512, 1024, 2048, 4096$. The dashed line has a slope of unity. Note the blue markers ($\phi = 0.6$) overlap with the green ones.*

### 3.2 Collective structure factors and real space contact values in copolymer systems

As discussed in the Section 2.3, for copolymers there are two contributions to the COM diffusion constant, the short time and length scale local segmental collision part and the long time and length scale microdomain-scale contribution. For a typical A/B copolymer system investigated here, in our previous work [33] we have shown that there exists a Lifshitz-like point at a copolymer packing fraction $\phi_L$, beyond which macroscopic phase separation becomes unstable and a continuous (often sharp) crossover occurs to a microemulsion-like locally clustered state as the B-B attraction in thermal energy units increases (Fig.3 (a)). In this microemulsion-like regime, the static structure factors never diverge at $k = 0$, but rather $S_{BB}(k)$ develops an intense finite wavevector microdomain scale peak at a characteristic length scale $k^*$ that grows with cooling or densification. This structural feature signals the emergence of strong local clustering and a correlated fluctuating microphase/microemulsion structure. We found that the value of $\phi_L$ generally falls in the range of $0.2 < \phi < 0.4$, suggesting the relevance of the dynamical weak-caging regime. Due to the potential importance of a strongly clustered state to the biomolecular condensate problem and our interest in understanding the role of clustering in the growth of friction, we focus our investigation solely for systems beyond the Lifshitz point, $\phi > \phi_L$.

The adopted A/B copolymer model is of an asymmetric interaction form with an attractive exponential potential only between B sites of contact strength $\epsilon_{BB}$ and with the spatial range parameter set to be $\alpha = 0.5$. For our initial investigation in this article, we focus on regular sequence multiblock copolymers of equal composition, $f_A = f_B = 1/2$. The block length is defined $M$, and the chain length is $N$.

To determine a characteristic temperature for the microemulsion crossover boundary, we adopt a thermodynamic and local structural clustering real space-based approach by calculating the dimensionless cohesive energy $\widehat{U}_{BB}$ of the system as a function $\epsilon_{BB}$, defined as,

$$\widehat{U}_{BB} = \frac{1}{2}\frac{\rho_B}{\epsilon_{BB}} \int_0^\infty dr\, 4\pi r^2 g_{BB}(r) u_{BB}(r) \qquad (15)$$

We consider the negative derivative of the dimensionless cohesive energy, $-d\widehat{U}_{BB}/d\epsilon_{BB}$, as a local metric of the crossover to the strongly clustered microemulsion state, and use the position of its peak as the transition point $\epsilon_{BB}^*$; see Fig.3(b). We previously showed [33] that this crossover temperature determined by such a thermodynamics-based procedure nicely quantitatively agrees with the classic linear extrapolation of $1/S_{BB}(k^*)$ to zero approach adopted in coarse grained mean-field theories [36,86] and an analogous analysis using PRISM theory [59–61,65,87].

Given our approach to predicting the microemulsion crossover boundary, examples of the general behavior of the BB radial distribution function and structure factors are shown in Figs. 3(c) and (d) at the microemulsion crossover boundary, $\epsilon_{BB}^*$. From Fig. 3(c) one sees that the B-B partial collective structure factors develops a low-wavevector peak (no divergence), indicating microphase clustering on the length scale of $1/k^*$ where $k^*$ is the position of the low-wavevector peak. The position and magnitude of B-B partial structure factor peak depends on block size $M$. Note that the local cage packing peak at $k\sigma \sim 6$ display no prominent changes at $\epsilon_{BB}/\epsilon_{BB}^* = 1$, suggesting the absence of any local strong caging effect, consistent with no locally driven glassy or gel-like dynamics. Fig. 3(d) presents the corresponding results for the segmental-scale local contact value between stickers, $g_{BB}(r = \sigma)$ at $\epsilon_{BB} = \epsilon^*$ which also depends on block size $M$. Results for the A-A and A-B partial structure factors and radial distribution functions are shown in Figure 1 of the SI, and display only weak changes with a decrease of temperature.

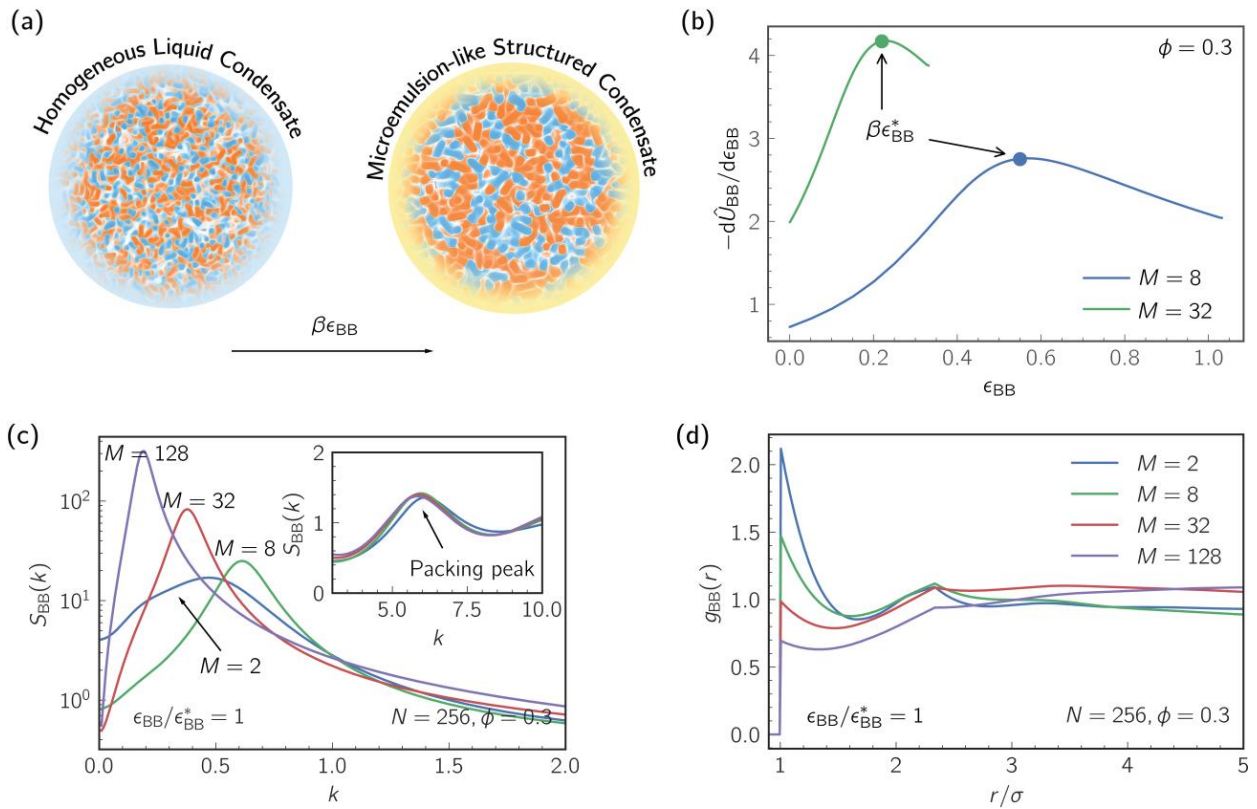

*Figure 3: (a) Illustration of the homogeneous microemulsion-like structure as the B-B attraction increases. Orange and cyan colors represent A and B sites, respectively. (b) Negative derivative*

of dimensionless cohesive energy, $-d\hat{U}_{BB}/d\epsilon_{BB}$. The characteristic peak is used as the metric for crossover to the strongly clustered state. $\phi = 0.3$. *(c)* B-B partial structure factor $S_{BB}(k)$ for different block sizes $M$ at the crossover boundary ($\epsilon_{BB}/\epsilon_{BB}^* = 1$). The inset shows the local cage or packing peak of $S_{BB}(k)$ around $k\sigma \sim 6$. The chain length is $N = 256$ and total copolymer volume fraction is $\phi = 0.3$. *(d)* B-B radial distribution function $g_{BB}(r)$ for different values of block size $M$ at the crossover boundary ($\epsilon_{BB}/\epsilon_{BB}^* = 1$). Note that $\phi > \phi_L$ in all cases of $M$.

We next quantify the dependence of the crossover $\epsilon_{BB}^*$ on the characteristic microdomain wavevector $k^*$ and block size $M$. Fig. 4(a) shows the strong dependence on $M$, with $\epsilon_{BB}^*$ decreasing rapidly with increasing block size. This strong dependence is also reflected locally in real space via the growth of the sticker-sticker number of nearest neighbors $n_{BB} = 4\pi\rho_B \int_\sigma^d dr\, r^2 g_{BB}(r)$ ($d$ is the position of first local minimum of $g_{BB}(r)$) as $\epsilon_{BB}$ increases. Fig.4(b) shows the growth of $n_{BB}$ for large $M$ is much slower than for smaller $M$, consistent with Fig.3(d) for the behavior of the sticker-sticker contact value. Thus, $\epsilon_{BB}^*$ is not determined by the absolute degree of clustering, but rather its rate of change with cooling.

Fig. 4(c) shows $k^*$ as the function of $M$ at $\epsilon_{BB}/\epsilon_{BB}^* = 1$. For $M > 32$, $k^* \propto 1/\sqrt{M}$, which translates to the microdomain size ($2\pi/k^*$) growing as the square root of $M$. This is an intuitive result since such scaling is proportional to the radius of gyration or end to end distance of a block for the compositionally symmetric copolymer model studied. For small $M < 32$, a large deviation from this simple scaling emerges, as expected, and $k^*$ varies non-monotonically for $2 < M < 16$.

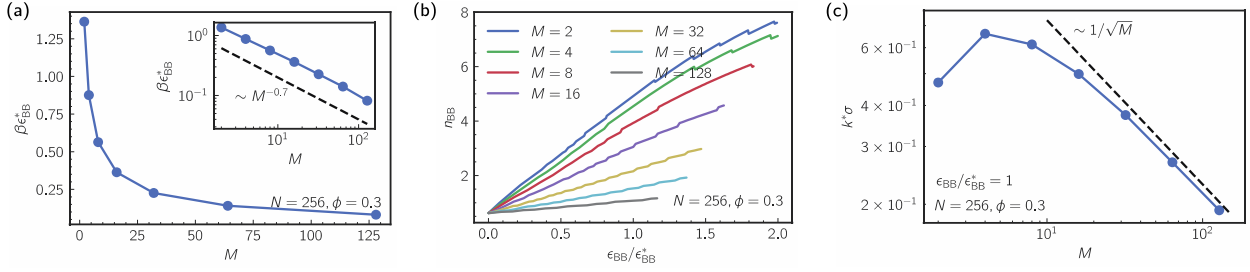

*Figure 4: (a) Crossover sticker-sticker attraction strength, $\epsilon_{BB}$, as a function of block size $M$. Here $N = 256$ and $\phi = 0.3$ for all subfigures shown. Inset shows the same data on a log-log scale. Note that $\epsilon_{BB}^*$ follows a $M^{-0.7}$ scaling for the whole range of $M$ investigated. (b) Number of nearest B-B (sticker-sticker) neighbors $n_{BB}$ as a function of normalized $\epsilon_{BB}$. (c) Characteristic wavevector $k^*$ as a function of $M$ at the crossover boundary $\epsilon_{BB}/\epsilon_{BB}^* = 1$. The dashed line indicates the scaling of $k^* \propto 1/\sqrt{M}$, i.e., the microdomain size scales as square root of block size.*

## 4. Copolymer CM diffusion constant and its connection with structure and sequence

### 4.1 Growth of total center-of-mass friction coefficient

We now present our main new dynamical results. We note in passing that SI Figure 2 shows that the per-monomer collective friction normalized by the elementary local segmental friction, $\Delta\zeta/\zeta_s$, for multiblock copolymer system under athermal conditions indeed properly reduces to that for homopolymer reference system. The growth of total center-of-mass friction constant $\zeta_{CM}$ for multiblock copolymer fluids as function of absolute values of $\epsilon_{BB}$, and also as a function of normalized $\epsilon_{BB}/\epsilon_{BB}^*$, are shown in Fig. 5. The $\zeta_{CM}$ is reported in a normalization form relative to the corresponding athermal homopolymer reference fluid where $\epsilon_{BB} = 0$. Since the athermal homopolymer friction constant is independent of $M$, this normalization merely shifts the curves.

Fig. 5 shows that that the absolute value of $\epsilon_{BB}^*$ has a significant effect on the extent of growth of total CM friction. Specifically, Fig. 5(a) shows that at fixed values of B-B attraction, copolymers with a larger block size $M$ have a higher total COM friction coefficient. Note that the curves for small $M$ end much earlier. This is due to the fact that the values of $\epsilon_{BB}^*$ become much smaller for large block sizes $M$, reflecting their higher tendency to microphase separate or cluster. On the other hand, if normalized by $\epsilon_{BB}^*$, the total friction constant for smaller block sizes is larger at fixed value of $\epsilon_{BB}/\epsilon_{BB}^*$ (Fig.5(b)). This suggests that at a "fixed distance" to the microemulsion boundary, copolymers with a smaller block size have a higher total friction compared to the system with a larger block size. This would seem to be physically intuitive since smaller blocks implies stronger finite size coupling of structural correlations on the local scale (which nucleates the BB clustering) and the microdomain scale. The dashed curves in Fig.5 are for $\phi = 0.5$, demonstrating that the qualitative behavior of diffusion as a function $\epsilon_{BB}$ do not changes with respect to $\phi$ for $\phi > \phi_L$.

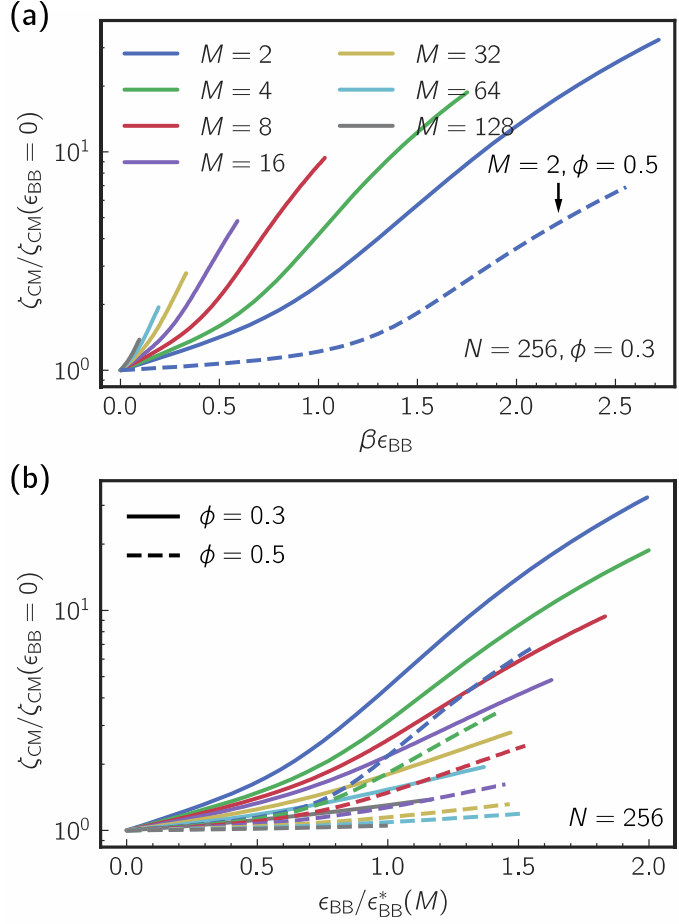

Figure 5: **(a)** Normalized total COM friction constant $\zeta_{CM}/\zeta_{CM}(\epsilon_{BB} = 0)$ as a function of absolute value of $\epsilon_{BB}$. Curves for different values of M are shown. One example for $M = 2$ and $\phi = 0.5$ is shown as the dashed curve. **(b)** Normalized total COM friction constant $\zeta_{CM}/\zeta_{CM}(\epsilon_{BB} = 0)$ as a function of normalized $\epsilon_{BB}/\epsilon_{BB}^*(M)$, where $\epsilon_{BB}^*(M)$ is the crossover transition to microemulsion which depends on M shown in Fig.4(a). $\phi = 0.3$ and $\phi = 0.5$ for different values of M are shown in solid and dashed lines, respectively.

Since the total COM friction constant has single-chain Rouse (diagonal or the self-component of the total force-total force correlation due to local collisional friction only) and collective (due to microdomain constraints) contributions, one can ask how much of the growth shown in Fig.5 is due to the collective friction. This motivates rewriting $\zeta_{CM}/\zeta_{CM}(\epsilon_{BB}=0)$ as,

$$\frac{\zeta_{CM}(\epsilon_{BB})}{\zeta_{CM}(\epsilon_{BB}=0)} = \frac{N\zeta_s(\epsilon_{BB})}{\zeta_{CM}(\epsilon_{BB}=0)} + \frac{N\Delta\zeta(\epsilon_{BB})}{\zeta_{CM}(\epsilon_{BB}=0)} \qquad (16)$$

where the first (second) term is the single-chain Rouse (collective) contribution.

Figs.6(a) and (b) compares the growth of the total COM friction and its single-chain Rouse contribution. For $M = 2$, Fig. 6(a) shows that collective contribution becomes dominant for $\epsilon_{BB} \gg \epsilon_{BB}^*$. On the hand, for the diblock copolymer $M = 128$, the collective contribution is only 20% to 40% of the total friction as a function of $\epsilon_{BB}$ (Fig. 6(b)). Combined with the results in Fig.5, we conclude the collective contribution to the total COM friction constant is much more important for copolymer systems with smaller block sizes. This is not a trivial deduction given the competing effects of reduction of the attractive dimensionless energy scale, but longer nature of the blocks, as $M$ grows.

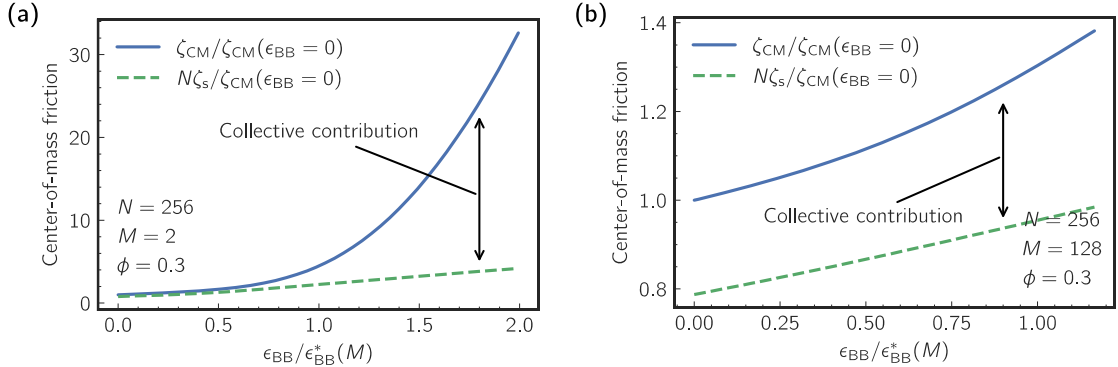

*Figure 6: **(a)** Comparison between the growth of the total COM friction, $\zeta_{CM}(\epsilon_{BB})/\zeta_{CM}(\epsilon_{BB} = 0)$, and the growth of single-chain Rouse contribution, $N\zeta_s/\zeta_{CM}(\epsilon_{BB} = 0)$. The difference between the two is the collective contribution. $N = 256, M = 2, \phi = 0.3$. **(b)** Diblock system. Comparison*

*between the growth of the total COM friction, $\zeta_{CM}(\epsilon_{BB})/\zeta_{CM}(\epsilon_{BB} = 0)$, and the growth of single-chain Rouse contribution, $N\zeta_s/\zeta_{CM}(\epsilon_{BB} = 0)$. The difference between the two is the collective contribution. $N = 256, M = 128, \phi = 0.3$.*

**4.2 Connection between local and domain-scale structures and the COM friction**

We now aim to understand the quantitative connection between the collective static structure and the growth of the total COM friction given by $\zeta_{CM} = N\zeta_s + N\Delta\zeta = N\zeta_s(1 + \Delta\zeta/\zeta_s)$. The change of $\zeta_{CM}$ comes from the local segmental friction coefficient $\zeta_s$ and the per-monomer collective friction $\Delta\zeta/\zeta_s$. Recall that $\zeta_s$ is proportional to the average contact value, $\zeta_s \propto f_A g_{AA}(\sigma) + f_B g_{BB}(\sigma) \equiv \overline{g(\sigma)}$, while $\Delta\zeta/\zeta_s$ is dependent on collective structural information contained in $\mathbf{C}(k)$, $\mathbf{S}(k)$, and $\boldsymbol{\omega}(k)$. We note that although $\Delta\zeta/\zeta_s$ is not *directly* connected to the contact value, physically is strongly correlated for causal reasons since microdomain formation is triggered by local clustering of like segments, as shown in our prior work [33]. Fig.7 shows that both the growth of average contact values and nondimensional per-monomer collective friction is much stronger for small block size compared to large block size, consistent with the results shown in Fig.5.

The average contact value, $\overline{g(\sigma)}$, quantifies the degree of very local clustering on the monomer scale, $\sigma$, and increases gradually over the entire range of $\epsilon_{BB}$ (Fig.7(a)). On the other hand, $\Delta\zeta/\zeta_s$ is directly connected to the formation of microdomains on the larger length scale $2\pi/k^*$. Fig.7(b) shows that $\Delta\zeta/\zeta_s$ only starts to grow significantly beyond $\epsilon_{BB}^*$ for small values of $M$, suggesting the growth of the low wavevector structure factor peak is the dominant origin of the growth of the collective friction, a physically sensible deduction.

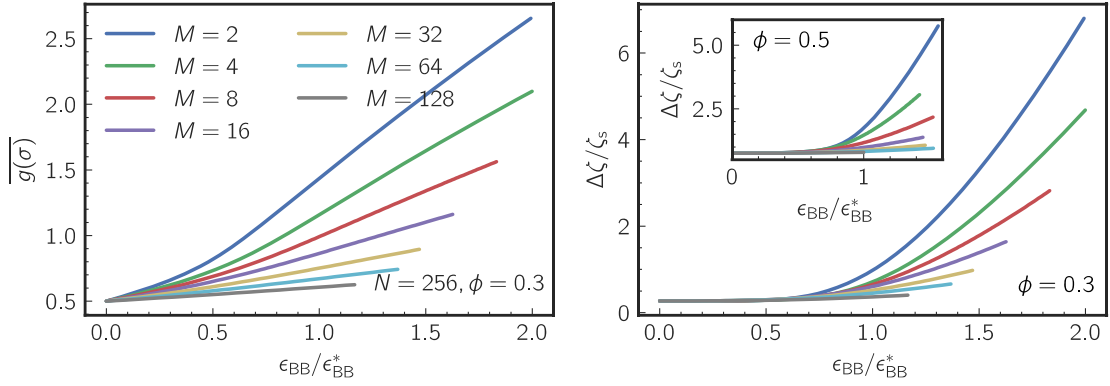

*Figure 7: **(a)** Average copolymer contact value $\overline{g(\sigma)}$ as a function of normalized attractive interaction $\epsilon_{BB}/\epsilon_{BB}^*$ for different values of block size $M$. **(b)** Non-dimensionalized per-monomer collective friction as a function of normalized attraction $\epsilon_{BB}/\epsilon_{BB}^*$. $N = 256$ and $\phi = 0.3$. The inset shows the results for $\phi = 0.5$.*

Next, we ask whether the slow growth of $\Delta\zeta/\zeta_s$ for large $M$ is due to the smaller microdomain scale structure factor peak. Fig.8(a) shows that the increase of the latter as a function of $\epsilon_{BB}/\epsilon_{BB}^*$ is strongest for larger $M$. Fig.8(b) shows $\Delta\zeta/\zeta_s$ as a function of microdomain structure factor peak height, and clearly demonstrates that even at the same value of $S_{BB}(k^*)$, the nondimensional per-monomer collective friction differs significantly for different values degrees of blockiness, $M$. Indeed, the ordering of the results in Fig.8(b) is the opposite of that in Fig.8(a). One can also ask whether the structure factor peaks normalized by the length scale of the microdomain for different values of $M$ collapse, i.e. does the intensity of the microdomain scale ordering scale quadratically with the measure of microdomain length scale? SI Figure 3(a) shows that $(k^*)^2 S_{BB}(k^*)$ does indeed tend to collapse for different values of $M$, albeit there are some deviations for very small values of $M = 2,4$. However, SI Figure 3(b) shows that nondimensional per-monomer collective friction for different values of $M$ do *not* collapse with $(k^*)^2 S_{BB}(k^*)$, similar to that in Fig.8(b).

The combined results shown in Fig.8 and SI Figure 3 show that $\Delta\zeta/\zeta_s$ is *not* simply linearly correlated to the microdomain scale peak height of the B-B partial structure factor nor its length-scale normalized version. Rather, it is causally connected to the direct correlation function, intramolecular correlation function, and all the partial structural factors that quantity the strength of the effective intermolecular mean square forces that enter the dynamic vertex and quantification of kinetic constraints in the theory.

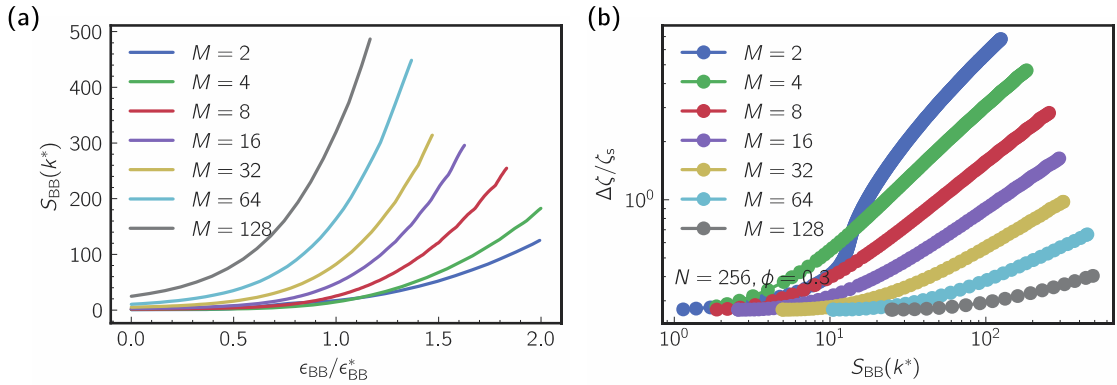

*Figure 8: **(a)** Growth of low wavevector structure BB factor peak, $S_{BB}(k^*)$, as a function of normalized attraction strength $\epsilon_{BB}/\epsilon_{BB}^*$. **(b)** Non-dimensional per-monomer collective friction, $\Delta\zeta/\zeta_s$, as a function of B-B structure factor microdomain peak height, $S_{BB}(k^*)$, where $N = 256$ and $\phi = 0.3$.*

To provide quantitative insight into how COM friction is connected to structure, we plot the dynamic vertex $V(k)$,

$$V(k) = \frac{1}{N}\int_0^\infty dt\, k^4 \sum_{\alpha,\gamma} \tilde{\omega}_{\alpha\gamma}(k,t)[\mathbf{C}\cdot\tilde{\mathbf{S}}(k,t)\cdot\mathbf{C}]_{\alpha\gamma} = \frac{k^4}{N}\sum_{\alpha,\gamma}\frac{A_{\alpha\gamma}G_{\alpha\gamma}}{\Gamma+\Lambda_I} + \frac{B_{\alpha\gamma}G_{\alpha\gamma}}{\Gamma+\Lambda_C} \qquad (17)$$

Fig.9(a) shows that with increasing $\epsilon_{BB}$, $V(k)$ is dominated by the low wavevector peak. Comparison with $S_{BB}(k^*)$ indicates clearly that the peak in the vertex originates from the growing microdomain scale structure factor intensity. We then ask whether $V(k)$ can be analytically simplified based on two approximations: 1) only consider the B-B terms in Eq. 3, which leads to $V_1(k) = \rho_B(N_B/N)k^2 C_{BB}^2(k)S_{BB}^2(k)\omega_{BB}^2(k)/(S_{BB}(k)+\omega_{BB}(k))$, and 2) an effective "homopolymer approximation", $V_2(k) = \rho k^2 \overline{C}^2(k) S^2(k) \omega^2(k)/(S(k)+\omega(k))$ where $\overline{C}(k) = (C_{AA}(k) + C_{BB}(k) + 2C_{AB}(k))/4$ and $S(k)$ is the total structural factor and $\omega(k)$ the total intramolecular correlation function. SI Figure 4 shows that these two approximations both capture the presence of the low wavevector peak of the vertex, and at roughly the correct positions. However, both give drastically *much larger values* compared to actual $V(k)$. Hence, these two approximations *cannot* be used to reliably capture the nondimensional per-monomer collective friction $\Delta\zeta/\zeta_s$.

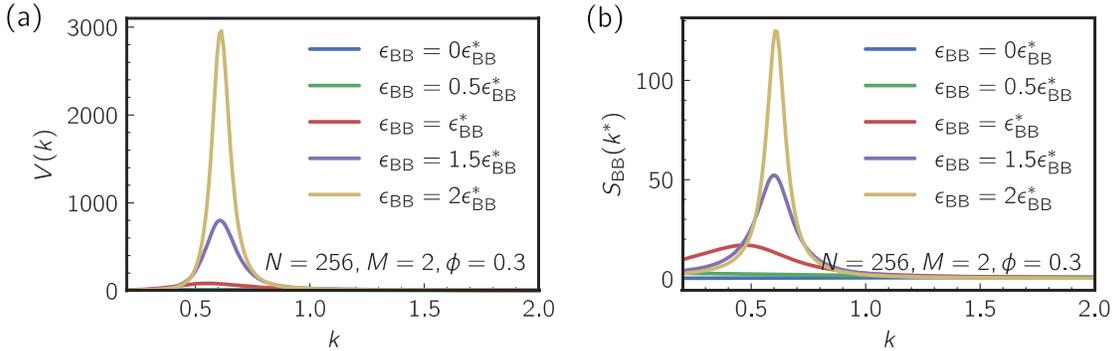

*Figure 9: (a) Dynamical vertex $V(k)$ defined in the text as a function of $k$ at various temperature. (b) The B-B partial collective structure factors. $N = 256, \phi = 0.3$.*

Even though $V_1(k)$ does not quantitatively approximate $V(k)$, the fact that it roughly indicates the correct position of the low wavevector peak in $V(k)$ prompts us to examine whether we can relate $\Delta\zeta/\zeta_s$ to $V_1(k)$. It is important to note that since $V_1(k) = \rho_B(N_B/N)k^2 C_{BB}^2(k)S_{BB}^2(k)\omega_{BB}^2(k)/(S_{BB}(k) + \omega_{BB}(k))$, under the assumption that $S_{BB}(k^*) \gg 1$ and $k^* \lesssim 1$, the maximum of $V_1(k)$ scales as $\propto (k^*)^2 C_{BB}^2(0)S_{BB}(k^*)$. Furthermore, with the reasonable simplification that the width of the sharp peak region of $S_{BB}(k)$, denoted as $\Delta k$, scales as $\Delta k \sim S_{BB}^{-1/2}(k^*)$, we obtain $\int dk V_1(k) \propto (k^*)^2 C_{BB}^2(0) S_{BB}^{1/2}(k^*)$. The above arguments suggest (as expected based on liquid state theory) that $k^* C_{BB}(k^*)$ acts as the effective force on the microdomain scale, while the factor of $S_{BB}^{1/2}(k^*)$ provides a measure of how effective forces on two tagged monomers are correlated on the microdomain scale. Hence, in total, this quantity is a strong candidate for capturing the strength of the effective mean square force exerted on a tagged copolymer by its surroundings due to microdomain formation and sticker-sticker attractive forces and repulsions.

The above analytic analysis inspires us to numerically investigate whether $\Delta\zeta/\zeta_s$ is linear in $(k^*)^2 C_{BB}^2(0) S_{BB}^{1/2}(k^*)$. Fig.10(a) indeed demonstrates that it is for three different values of $M$. However, the slope depends on $M$. This additional dependence on $M$ is incorporated in the full friction constant equation (Eq. (4) and Eq. (17)) and cannot be extracted from the approximation of $V_1(k)$. Hence, the relatively simple physical picture obtained from the physically motivated analytic analysis provides a solid understanding of our numerical results.

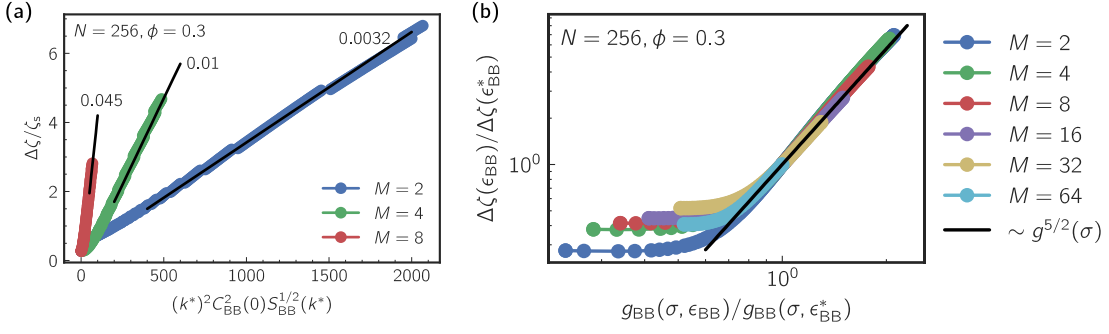

*Figure 10: **(a)** Nondimensional per-monomer collective friction, $\Delta\zeta/\zeta_s$, as a function of the estimate of the relevant mean square force slowing down of copolymer diffusion, $(k^*)^2 C_{BB}^2(0) S_{BB}^{1/2}(k^*)$. Only small values of M are chosen since their growth of friction is the most prominent. Black lines are the best linear fits with slopes indicated. **(b)** Per-monomer collective friction, normalized by its value at $\epsilon_{BB}^*$, as a function of B-B contact value normalized by its value at $\epsilon_{BB}^*$.*

We next ask how the collective friction is correlated with the *real space* structure, characterized by its most local metric on the scale the actual real bare forces are exerted, the contact value. Since only the B-B contact value increase significantly with $\epsilon_{BB}$, we choose $g_{BB}(\sigma)$ as the metric, which also quantifies the intensification of clustering and sticky monomer collisions. Fig.10(b) shows that if normalized by their values at the transition temperature, $\epsilon_{BB}^*$, all the theoretical data for $\Delta\zeta(\epsilon_{BB})/\Delta\zeta(\epsilon_{BB}^*)$ as a function of $g_{BB}(\sigma,\epsilon_{BB})/g_{BB}(\sigma,\epsilon_{BB}^*)$ remarkably collapses for a wide range of $M$ values. This establishes that the per-monomer collective friction $\Delta\zeta(\epsilon_{BB})$ is predicted to grow with $g_{BB}(\sigma)$ as an apparent power law. Empirical fitting suggests $\Delta\zeta(\epsilon_{BB}) \sim g_{BB}^{5/2}(\sigma)$, see Fig.10(b). We do emphasize this exponent may depend on the specific parameters and interaction model adopted, and should not be assumed to be universal. The correlation between $\Delta\zeta(\epsilon_{BB})$ and $g_{BB}$ is perhaps not surprising given the understanding previously developed from PRISM theory [33] that local clustering on the monomer scale is intrinsically and inevitably coupled with (or one can say nucleates) the formation of microdomains on the $1/k^*$ scale. As the B-B attraction increases, both local contacts between B sites and the magnitude of the low wavevector microdomain scale structuring increases. In addition, the direct correlation function $C_{BB}(k)$, and its integrated value $C_{BB}(k=0)$, which enter the theory as effective or renormalized interactions or forces, also change as $\epsilon_{BB}$ increases. However, precisely why $\Delta\zeta(\epsilon_{BB})$ scales with $g_{BB}(\sigma)$ of a power law is not a priori obvious.

**4.3 Correlation between copolymer blockiness and slowing down of COM diffusion**

Recent simulation studies have found that the sequence characteristics are a major determinant of the dynamics and material properties of biomolecular condensates [88]. Specifically, the authors studied model proteins that consist of 50% negatively charged glutamic acid (E) and 50% positively charged lysine (K) residues E-K variants [89]. They first determined the dense coexisting phase density by employing constant pressure simulation and subsequently perform Langevin dynamics and nonequilibrium MD (NEMD) simulations to obtain the diffusion constants and viscosity. The key results relevant to our study are (i) the equilibrium density of the dense phase strongly correlates with the polymer sequence blockiness parameter $\Omega$ at the fixed temperature, (ii) the polymer diffusion constants in the dense phase decrease as $\Omega$ increases at the fixed temperature, and (iii) the viscosity and diffusion constant are inversely linear correlated, suggesting a simple physical picture of effective "Rouse"-like dynamics (see Section 5). The latter parameter is defined following [90–92],

$$\text{SCD} = \frac{1}{n}\sum_{i<j} \sigma_i\, \sigma_j |i-j|^{1/2} \qquad (18)$$

where $\sigma_i = 1$ if the type of segment is A, and $\sigma_i = -1$ if the type of segment is B. Since the range of the variable SCD also depends on the copolymer composition and polymer chain length, we define a normalized SCD as,

$$\Omega = \frac{\max(\text{SCD}) - \text{SCD}}{\max(\text{SCD}) - \min(\text{SCD})} \quad (19)$$

This constrains the values of $\Omega$ to lie between 0 and 1, where $\Omega = 1$ corresponds to the most blocky sequence (the diblock), and $\Omega = 0$ corresponds to the least blocky sequence which for our 50/50 composition models is the $M = 2$ multiblock.

The comparison between different sequences in Ref.[88] was made at fixed temperature, corresponding to fixed $\epsilon_{BB}$ in our model. However, as shown in Fig.5, the range of $\epsilon_{BB}$ accessible is greatly limited by the value of the crossover boundary $\epsilon_{BB}^*$ which decrease as $M$ grows. In order to compare the total COM friction at a fixed $\epsilon_{BB}$, e.g. $\beta\epsilon_{BB} = 0.5$, we seek to find empirical fit equations of our theoretical $\zeta_{CM}(\epsilon_{BB})$ data for different values of $M$, and thereby estimate the friction at a given $\epsilon_{BB}$ via extrapolation. Since $\zeta_{CM} = N\zeta_s + N\Delta\zeta$, we first study the scaling behavior of $\zeta_s - \zeta_s(\epsilon_{BB} = 0)$ and $\Delta\zeta - \Delta\zeta(\epsilon_{BB} = 0)$ as functions of $\epsilon_{BB}$.

Fig.11a shows that for both small and large values of $\epsilon_{BB}$, $\zeta_s - \zeta_s(\epsilon_{BB} = 0)$ scales approximately linearly in $\epsilon_{BB}$ for all values of $M$. There are crossover regions for intermediate values of $\epsilon_{BB}$. However, since we are interested in the deduced behavior at large $\epsilon_{BB}$, such a crossover region is unimportant to our analysis. Next, we interestingly find that $\Delta\zeta - \Delta\zeta(\epsilon_{BB} = 0)$ scales with a 7/2 power of $\epsilon_{BB}$ for $\epsilon_{BB} > \epsilon_{BB}^*$ (Fig.11b). Per our results above, a 7/2 exponent is naturally interpreted as arising from one power of the scaling of the short time friction constant $\zeta_s$ and via a 5/2 power from the contribution of microdomain scale structure factors and the direct correlation function. As a result, we can write for the total COM friction constant as $\zeta_{CM} - \zeta_{CM}(\epsilon_{BB} = 0) \approx A\epsilon_{BB} + B\epsilon_{BB}^{7/2}$, where $\zeta_{CM}(\epsilon_{BB} = 0)$ is the reference homopolymer friction constant. Fig.11c shows the fitted curves to the numerical results. For $\epsilon_{BB}$ outside of the accessible range of the PRISM theory calculation, we use the fitted curve $A\epsilon_{BB} + B\epsilon_{BB}^{7/2}$ to infer $\zeta_{CM}$.

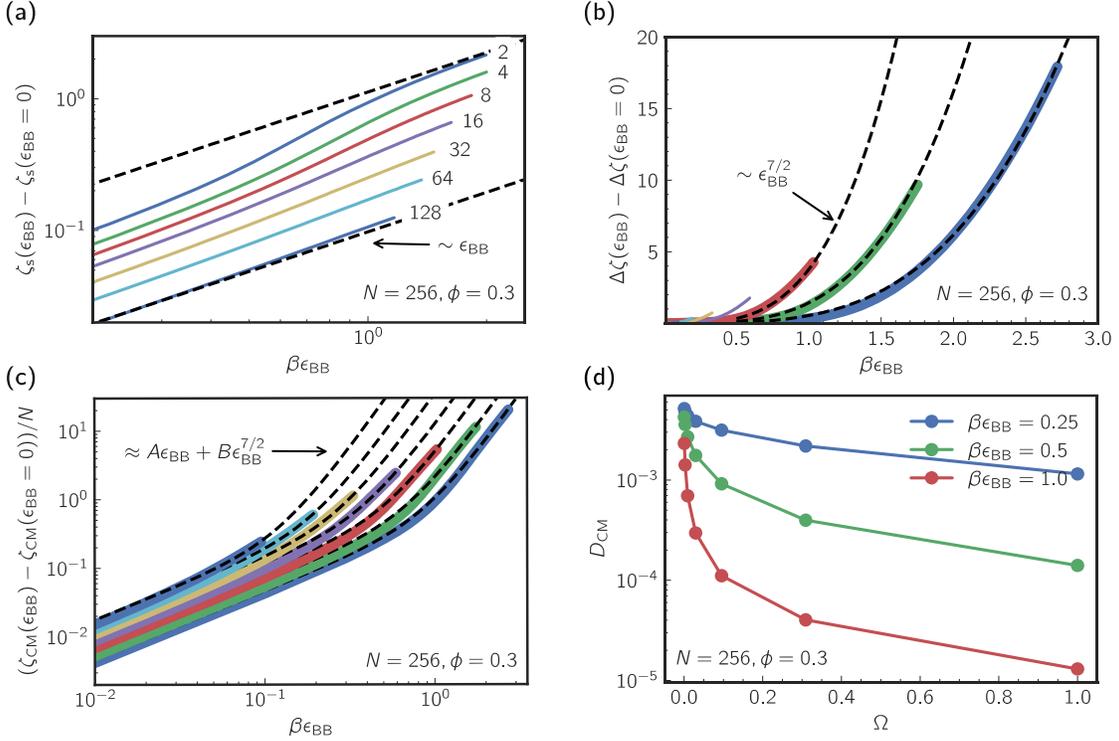

Figure 11: (a) Power-law scaling of $\zeta_s - \zeta_s(\epsilon_{BB} = 0)$ for $\phi = 0.3$. Dashed lines are guides to the eye with a slope of unity. The number next to each curve indicates the value of block size M. (b) Power-law scaling of $\Delta\zeta - \Delta\zeta(\epsilon_{BB} = 0)$ for $\phi = 0.3$. Dashed curves are guides to the eye of the form $\sim \epsilon_{BB}^{7/2}$. (c) Empirical fitting to $\zeta_{CM} - \zeta_{CM}(\epsilon_{BB} = 0)$ using $A\epsilon_{BB} + B\epsilon_{BB}^{7/2}$ with numerical fit prefactor parameters A and B. Note that values of A and B depends on M, $\phi$ and N. (d) Center-of-mass diffusion constant $D_{CM}$ as a function sequence blockiness parameter $\Omega$ for different values

*of $\epsilon_{BB}$. $\Omega$ is calculated for a system with different $M$. Note that $M = 128$ is the diblock system, which has $\Omega = 1$ and $M = 2$ and which has the smallest value of $\Omega$.*

Using the so-obtained function $A\epsilon_{BB} + B\epsilon_{BB}^{7/2}$ for each value of $M$, we calculate the inferred value of total COM friction constant $\zeta_{CM}$ at a given value of $\epsilon_{BB}$ and corresponding diffusion constant $D_{CM} = k_B T/\zeta_{CM}$. Fig.11(d) shows how the COM diffusion constant monotonically decreases with sequence blockiness as the metric parameter $\Omega$ increases. Note that $M = 2$ has the smallest value of $\Omega$ and $M = 128$ (diblock) has $\Omega = 1$. For larger value of $\epsilon_{BB}$, the extent of reduction of $D_{CM}$ as $\Omega$ grows from 0 to 1 increases. Interestingly, the dependence of the diffusion constant on $\Omega$ is qualitatively similar to that shown in the simulation study of Ref. [88]. Compared to the simulation results, our theoretical prediction does capture the general qualitative behavior of $D_{CM}(\Omega)$ where it decreases relatively fast for small $\Omega$ and then tends to level off for larger $\Omega$. However, we again note that in the simulation study [88], the diffusion constant was measured in the dense phase which has a polymer density that monotonically increases by a factor of roughly 1.5-2 as $\Omega$ varies between 0 and 1. In contrast, each of our curves shown in Fig. 11(d) are at fixed copolymer volume fraction $\phi$. Therefore, our theoretical results in Fig.11d should not be quantitatively compared to the simulation study [88]. But based on our theoretical work we propose that understanding the sequence-dependent dynamics of copolymer and condensate systems requires distinguishing between effects originating from polymer volume fraction and temperature (which is linked to the strength of attraction). We anticipate that this insight will guide future simulation studies to accurately disentangle the dynamical consequences of these two variables. This will hopefully pave the way to a better understanding of how the sequence influences the diffusion slowdown in copolymer and condensate liquids.

**4.4 Comparison to dynamical results based on PRISM-RPA theory of structure**

We now briefly explore the comparison between our theoretical friction constant calculations of using full PRISM theory structural input that includes correlated concentration fluctuations per a polymeric microemulsion and does not exhibit any mean field microphase spinodals, with analogous dynamical predictions based on using the mean field PRISM-RPA theory input [33,54]. Recall that PRISM-RPA theory does predicts a strict microphase separation (MiPS) transition at a certain critical $\epsilon_{BB}^*$ at which all the partial collective static structural factors diverge at $k^*$. Hence, it is straightforward to appreciate that the friction constant must also diverges at $\epsilon_{BB}^*$ due to the diverging structure factors in the integral of Eq. 2. Indeed, we find that $\Delta\zeta/\Delta\zeta(\epsilon_{BB} = 0)$ diverges at $\epsilon_{BB}^*$ for our multiblock system (Fig.12a), and as a power law in the difference variable $\epsilon_{BB}^* - \epsilon_{BB}$ (Fig.12b), consistent with classic mean field critical-like behavior.

Beyond the spurious divergence, in contrast to our dynamical predictions based on the full PRISM theory structural input, the corresponding CM diffusion constant results using PRISM-RPA theory input shows little block size ($M$) dependence of friction constant growth other than its effect on $\epsilon_{BB}^*$. Fig.12b shows explicitly that the growth of friction in the vicinity of the spinodal is weakly dependent on $M$. We note that this is vastly different from the results obtained using the full PRISM theory structural input where the growth of friction does not collapse based on the rescaled $\epsilon_{BB}/\epsilon_{BB}^*$ (see Fig.7b). We thus conclude that comparison between the full PRISM and PRISM-RPA theories for structural input to the dynamical theory shows the latter mean field approach predicts strong and sudden growth of friction as $\epsilon_{BB}$ approaches to $\epsilon_{BB}^*$, in qualitative disagreement with the gradual growth of friction in the full PRISM theory based predictions (Fig.12c).

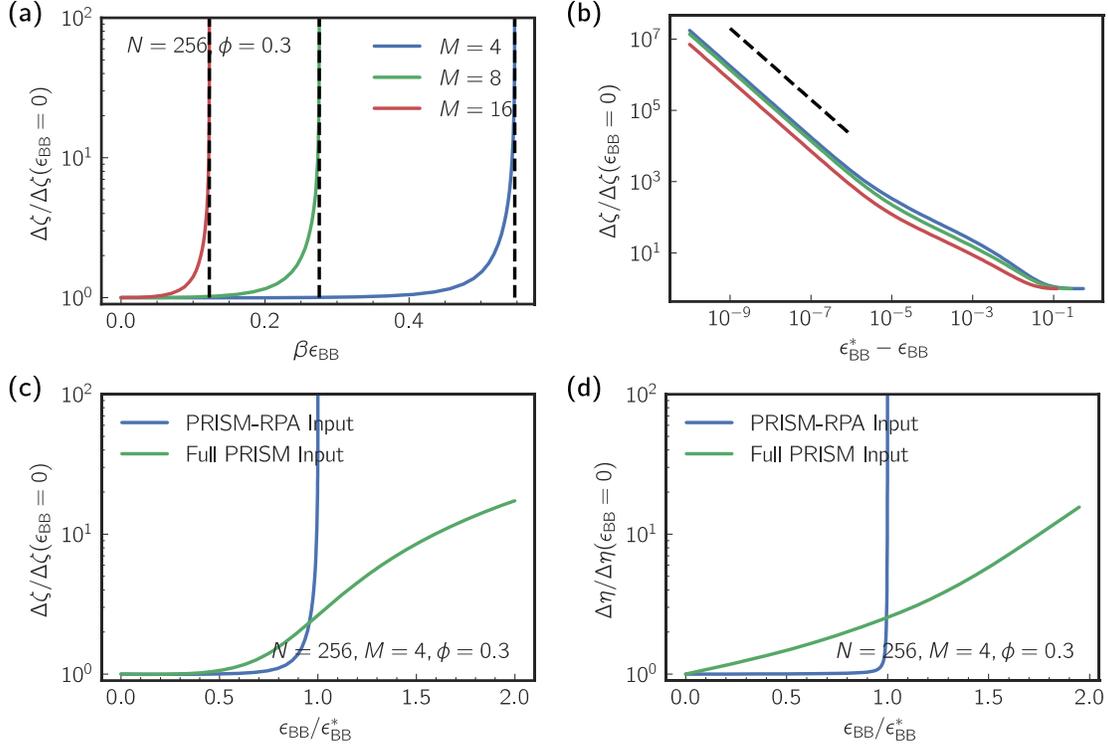

*Figure 12: PRISM-RPA theory based dynamical results for the growth of COM friction (a) Normalized collective friction, $\Delta\zeta/\Delta\zeta(\epsilon_{BB}=0)$, as a function of $\epsilon_{BB}$ for multiblock systems with $\phi = 0.3$. (b) The power-law critical scaling of friction in the vicinity of spinodal where $\epsilon_{BB}^*$ is the reduced energy scale at which the structure factor diverges. (c) Comparison between calculation of friction using full PRISM theory structural input versus using PRISM-RPA theory input to our dynamical theory for a representative example of $M = 4, \phi = 0.3$. (d) Same as panel (c) but for the collective friction $\Delta\zeta$.*

## 5. Viscosity in multiblock copolymer system

The shear viscosity and self-diffusion constant are distinct dynamical facets of any liquid. For polymers, the Rouse model [53] postulates a straightforward linear correlation between viscosity, denoted by $\eta$, and the cumulative friction experienced by a single chain, represented by $N\zeta_s$, which is generally well behaved for homopolymer melts. When $N\zeta_s$ is substituted with $\zeta_{CM}$ from our copolymer theory, it results in an effective Rouse prediction of a single-chain contribution to the total viscosity of a copolymer system, referred to as $\eta_R$ (see Eq. (10)). While $\zeta_{CM}$ is calculated via the interchain static pair correlation, $\eta_R$ primarily accounts for the single-chain contribution and does not explicitly consider inter-chain forces and stress.

As discussed in Section 2.4, we have analyzed the role of collective viscosity, $\Delta\eta$. Fig.13 shows representative calculations of the ratio of the collective to total shear viscosity, $\Delta\eta/\eta$. The results indicate that for copolymers with small block size, the contribution of collective viscosity is relatively minor ($\Delta\eta/\eta < 0.1$). This result is in stark contrast with dynamical predictions based using the mean-field PRISM-RPA theory as structural input. Recall Fig. 12(d) shows that if using the static correlation information obtained from PRISM-RPA mean field theory, the collective viscosity literal diverges at the spinodal $\epsilon_{BB}^*$, in contrast with the gradual growth predicted using full PRISM theory calculated correlation functions that includes stabilizing fluctuation effects. Fig.13 also shows that as the block size $M$ and $\epsilon_{BB}$ increase, the role of $\Delta\eta$ becomes increasingly prominent, manifested as the apparent upturn trends. Overall, our results suggest that over the range of temperatures accessible to PRISM theory and (we believe) likely relevant to biological systems (including well into the polymeric microemulsion state), the collective friction is perturbative compared to the single-chain entropic contribution to the total viscosity. The obvious caveat that even deeper in the microemulsion "phase" the collective contribution might matter since we predict that it does grow more strongly with cooling than the single chain entropic viscosity.

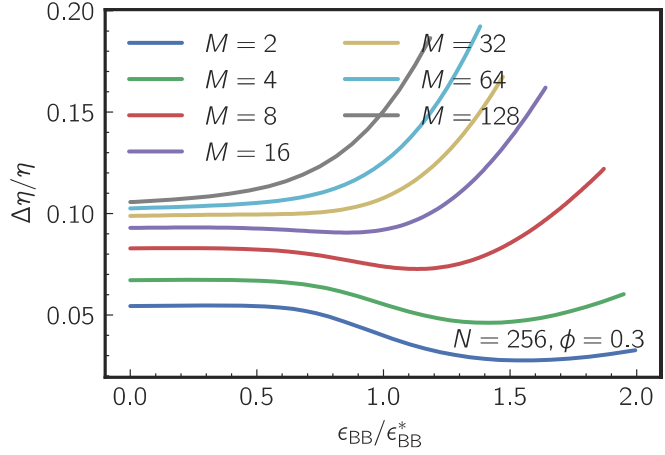

*Figure 13: Relative contribution of the collective viscosity to the total viscosity, $\Delta\eta/\eta$, as a function of normalized sticker-sticker attraction $\epsilon_{BB}/\epsilon_{BB}^*$ for different values of block size M for $N = 256$ and $\phi = 0.3$.*

## 6. Conclusions and Discussion

We have proposed a general microscopic dynamical theory for the center-of-mass diffusion constant and shear viscosity of concentrated A/B copolymer solutions and melts in the so-called weak caging regime. As a first application, the theory was implemented for regular multiblock copolymers of highly variable block length at fixed composition, systems relevant to synthetic macromolecular science. Our approach employs equilibrium correlation functions computed from PRISM theory as input to quantify length-scale-dependent interchain dynamical constraints in order to predict the COM friction coefficient. Our findings elucidate the relationship between the copolymer block size ($M$), the normalized sticker-sticker attraction ($\epsilon_{BB}/\epsilon_{BB}^*$) that drives local clustering and fluctuating microdomain formation, and the total COM friction coefficient ($\zeta_{CM}$). At fixed $\epsilon_{BB}/\epsilon_{BB}^*$, we predict that copolymers with smaller block sizes exhibit a more pronounced growth of total CM friction and that the collective friction contribution associated with microdomain formation is dominant. We further demonstrated that the collective contribution ranges from 20% to 40% of the total friction constant renormalization for a diblock copolymer with $M = 128$.

At a more detailed level, our study reveals a complex, yet understandable, connection between the degree of fluctuating microdomain order and the growth of the total COM friction in multiblock copolymer fluids. This has been achieved by dissecting $\zeta_{CM}$ into its two primary components: increase of the local segmental friction coefficient $\zeta_s$ due to monomer scale physical clustering as encoded in the contact value of the interchain pair correlation function, and the non-dimensionalized per-monomer collective friction contribution $\Delta\zeta/\zeta_s$ associated with microdomain formation on a larger length scale. We find that both these contributions grow strongly with cooling or increase of sticky monomer attraction for smaller block sizes. We have also demonstrated a nuanced relationship exists between $\Delta\zeta/\zeta_s$, $S(k)$, and $C(k)$, emphasizing that $\Delta\zeta/\zeta_s$ is not simply linearly correlated to the microdomain scale peak height, nor its length-scale normalized version, nor the magnitude of the bare sticky monomer attraction energy. However, we have identified an underlying simplicity associated with an apparent power-law correlation between $\Delta\zeta(\epsilon_{BB})$ and $g_{BB}^{5/2}(\sigma)$, though we emphasize that there is no a priori reason that this exponent is universal. This correlation of short-range structure and friction enhancement underscores the causal interdependence between the local real space contacts between B sites, the intensity of the low wavevector microdomain structuring, and the renormalized interchain interactions (beyond an empirical chi-parameter in a mean field framework [36,86]) encoded in the site-site direct correlation functions. These findings more generally also highlight the complex interplay between different length scales and correlated structure on friction and copolymer diffusion.

We have also investigated the copolymer sequence dependence of friction, establishing a clear correlation between the statistical metric Ω employed in recent biocondensate studies which quantifies sequence blockiness [92], and the COM diffusion constant at fixed reduced temperature and volume fraction. Our findings suggest that care should be taken when dissecting the pure sequence-dependent dynamical property in experiments. For example, in the biomolecular condensate context, the concentration of the polymer density of the dense coexisting condensate phase is sequence-dependent which may crucially affect dynamics. Finally, the role of collective stress contribution associated with emergent microdomain formation to the shear viscosity was assessed, and found to provide a relatively minor contribution compared to the classic single chain (Rouse model) contribution. This results in a near linear relationship between viscosity ($\eta$) and the total center-of-mass friction ($\zeta_{CM}$) enhancements in the copolymer systems studied.

Concerning the basic statistical mechanical formulation and approximations of our theory, we have assumed the systems are in a "weak caging" regime which in the language of glass physics corresponds being well removed from the ideal mode-coupling dynamical arrest transition which in reality signals a crossover to transient localization and activated dynamics [93]. This simplification is a priori verified to be justified by the relatively modest frictional enhancements over the reduced temperature range studied. Specifically, frictional enhancements are less than of order 30 or so, a magnitude consistent with the amount of slowing down that a weak caging approach accurately captures for the self-diffusion constant and viscosity of the foundational hard sphere colloidal suspension [50–52]. To further buttress this perspective, we note that the maximum sticker-sticker attraction strength investigated is $\epsilon_{BB} \approx 2.7 k_B T$, which is smaller than the ideal MCT transition ($\sim 3 - 4 k_B T$) due to physical bond formation and gelation predicted for single-component and biphasic mixture colloid systems [40]. The range and energy scale considered in this study is relevant to van der Waals and some hydrophobic interactions [94,95]. For other relevant attractions such as $\pi - \pi$ stacking and hydrogen bonding which have much shorter range and stronger association energy [94,95], the "weak caging" regime may become less accurate and physical gelation signaled by an ideal MCT transition is possible.

The fraction of stickers will also impact how close the system is to a dynamical ideal MCT transition or crossover. For copolymers with a much smaller fraction of stickers (e,g,, so-called associating polymers [96–98]) compared to what is studied here (50%), higher sticker-sticker attractions would presumably be required to achieve the same degree of growth of structural factors. This requirement is particular relevant to potential physical gel formation [99], and calls for a more direct treatment of strong *attractive forces*, a notion supported by previous theoretical work [80,81]. This may lead to a stronger growth of friction with cooling into a fluctuating structural state than that obtained here. Importantly, a more accurate and explicit treatment of the strong attractive forces beyond their consequences on pair packing structure [80,81] (entering only via the dynamical vertex) may have implications for the relevance of an ideal MCT dynamical arrest transition, which might serve as the necessary condition for the non-equilibrium aging observed in certain biomolecular condensate systems [5,30,31] and very recent computer simulation and theoretical studies of such phenomena [100–102]. Another distinct type of caveat is that the validity of the weak caging dynamical approximation may be sensitive to the specific choice of interaction potential model adopted. In our present study, we adopted a minimalist model where the spacer-spacer and spacer-sticker interactions are purely repulsive, and only the sticker-sticker interaction is attractive. It is possible that if the spacer-spacer and/or spacer-sticker interactions were weakly attractive, stronger sticker-sticker interactions could be needed to achieve the same degree of local clustering and microdomain scale ordering, which might push the system closer to the mode-coupling crossover and into a regime where physical gelation and activated bond breaking processes become important. Future theoretical studies in these directions are planned for synthetic A/B multiblock copolymers and aperiodic biomolecular condensates.

Our theoretical results can be tested based on systematic experiments and/or simulations on synthetic multiblock copolymers which are microstructured but globally homogeneous (polymeric microemulsion). We suggest that our approach has the potential to open up a new perspective on biomolecular condensates of variable and generally aperiodic sequences that we previously proposed can organize into a globally disordered fluctuating microemulsion state upon phase separation [33]. Of course, our present analysis of model copolymers with different regular block sizes and only sticky-sticky group attractions does not address the full complexity of real biomolecules. This presents another potential area for further exploration, notably the detailed investigation of sequence and interaction effects on dynamical slowing, which we believe is best pursued in the context of specific biomolecular condensate systems.

**Acknowledgments.** This work was supported by the Morris Professorship held by K.S.S. at UIUC.

**References.**

101  S. Blazquez, I. Sanchez-Burgos, J. Ramirez, T. Higginbotham, M. M. Conde, R. Collepardo-Guevara, A. R. Tejedor and J. R. Espinosa, *Advanced Science*, 2023, 2207742.

102  R. Takaki, L. Jawerth, M. Popović and F. Jülicher, *PRX Life*, 2023, **1**, 013006.